1

# Modern Classical Electrodynamics and Electromagnetic Radiation - Vacuum Field Theory Aspects

Nikolai N. Bogolubov (Jr.)[1], Anatolij K. Prykarpatski[2]
[1]V.A. Steklov Mathematical Institute of RAS, Moscow
[2] The AGH University of Science and Technology, Krakow
[2]Ivan Franko State Pedagogical University, Drohobych, Lviv region
[1]Russian Federation
[2]Poland
[2]Ukraine

## 1. Introduction

*"A physicist needs his equations should be mathematically sound and that in working with his equations he should not neglect quantities unless they are small"*
P.A. M. Dirac

Classical electrodynamics is nowadays considered (29; 57; 80) the most fundamental physical theory, largely owing to the depth of its theoretical foundations and wealth of experimental verifications. Electrodynamics is essentially characterized by its Lorentz invariance from a theoretical perspective, and this very important property has had a revolutionary influence (29; 57; 73; 80; 110) on the whole development of physics. In spite of the breadth and depth of theoretical understanding of electromagnetism, there remain several fundamental open problems and gaps in comprehension related to the true physical nature of Maxwell's theory when it comes to describing electromagnetic waves as quantum photons in a vacuum: These start with the difficulties in constructing a successful Lagrangian approach to classical electrodynamics that is free of the Dirac-Fock-Podolsky inconsistency (53; 110; 111), and end with the problem of devising its true quantization theory without such artificial constructions as a Fock space with "indefinite" metrics, the Lorentz condition on "average", and regularized "infinities"  (73) of S-matrices. Moreover, there are the related problems of obtaining a complete description of the structure of a vacuum medium carrying the electromagnetic waves and deriving a theoretically and physically valid Lorentz force expression for a moving charged point particle interacting with and external electromagnetic field. To describe the essence of these problems, let us begin with the classical Lorentz force expression

$$F := eE + eu \times B, \qquad (1.1)$$

where $e \in \mathbb{R}$ is a particle electric charge, $u \in T(\mathbb{R}^3)$ is its velocity vector, expressed here in the light speed $c$ units,

$$E := -\partial A/\partial t - \nabla \varphi \tag{1.2}$$

is the corresponding external electric field and

$$B := \nabla \times A \tag{1.3}$$

is the corresponding external magnetic field, acting on the charged particle, expressed in terms of suitable vector $A : M^4 \to \mathbb{E}^3$ and scalar $\varphi : M^4 \to \mathbb{R}$ potentials. Here "$\nabla$" is the standard gradient operator with respect to the spatial variable $r \in \mathbb{E}^3$, "$\times$" is the usual vector product in three-dimensional Euclidean vector space $\mathbb{E}^3$, which is naturally endowed with the classical scalar product $< \cdot, \cdot >$. These potentials are defined on the Minkowski space $M^4 \simeq \mathbb{R} \times \mathbb{E}^3$, which models a chosen laboratory reference frame $\mathcal{K}$. Now, it is a well-known fact (56; 57; 70; 80) that the force expression (1.1) does not take into account the dual influence of the charged particle on the electromagnetic field and should be considered valid only if the particle charge $e \to 0$. This also means that expression (1.1) cannot be used for studying the interaction between two different moving charged point particles, as was pedagogically demonstrated in (57).

Other questionable inferences, which strongly motivated the analysis in this work, are related both to an alternative interpretation of the well-known *Lorentz condition*, imposed on the four-vector of electromagnetic potential $(\varphi, A) \in T^*(M^4)$ and the classical Lagrangian formulation (57) of charged particle dynamics under an external electromagnetic field. The Lagrangian approach is strongly dependent on the important Einsteinian notion of the rest reference frame $\mathcal{K}_\tau$ and the related least action principle, so before explaining it in more detail, we first analyze the classical Maxwell electromagnetic theory from a strictly dynamical point of view.

## 2. Relativistic electrodynamics models revisited: Lagrangian and Hamiltonian analysis

### 2.1 The Maxwell equations revisiting

Let us consider the additional Lorentz condition

$$\partial \varphi / \partial t + < \nabla, A > = 0, \tag{1.4}$$

imposed *a priori* on the four-vector of potentials $(\varphi, A) : M^4 \to \mathbb{R} \times \mathbb{E}^3$, which satisfy the Lorentz invariant wave field equations

$$\partial^2 \varphi / \partial t^2 - \nabla^2 \varphi = \rho \tag{1.5}$$

and

$$\partial^2 A / \partial t^2 - \nabla^2 A = J, \tag{1.6}$$

where $\rho : M^4 \to \mathbb{R}$ and $J : M^4 \to \mathbb{E}^3$ are, respectively, the charge and current densities of the ambient charged matter, which satisfy the charge continuity equation

$$\partial \rho / \partial t + < \nabla, J > = 0. \tag{1.7}$$

Based on equations (1.4), (1.5) and (1.7) one easily derives the classical electromagnetic Maxwell field equations (56; 57; 70; 80)

$$\nabla \times E + \partial B/\partial t = 0, \quad <\nabla, E> = \rho, \nabla \times B - \partial E/\partial t = J, \quad <\nabla, B> = 0,$$

for all $(t, r) \in M^4$ with respect to the chosen reference frame $\mathcal{K}$.

Notice here that Maxwell's equations (1.8) do not directly reduce, via definitions (1.2) and (1.3), to the wave field equations (1.5) without the Lorentz condition (1.4). This fact is very important, and suggests that when it comes to a choice of governing equations, it may be reasonable to replace Maxwell's equations (1.8) and (1.7) with the Lorentz condition (1.4), (1.5) and the continuity equation (1.7). From the assumptions formulated above, one infers the following result.

**Proposition 2.1.** *The Lorentz invariant wave equation (1.5) for the electric potential $\varphi : M^4 \to \mathbb{R}$ together with the Lorentz condition (1.4) and the charge continuity relationship (1.7) are completely equivalent to the Maxwell field equations (1.8).*

*Proof.* First of all it is easy to observe that equation (1.6) follows from (1.5) and (1.7). Substituting (1.4), into (1.5), one easily obtains

$$\partial^2 \varphi/\partial t^2 = - <\nabla, \partial A/\partial t> = <\nabla, \nabla \varphi> + \rho, \tag{1.8}$$

which implies the gradient expression

$$<\nabla, -\partial A/\partial t - \nabla \varphi> = \rho. \tag{1.9}$$

Taking into account the electric field definition (1.2), expression (1.9) reduces to

$$<\nabla, E> = \rho, \tag{1.10}$$

which is the second of the first pair of Maxwell's equations (1.8). Now upon applying $\nabla \times$ to definition (1.2), we find, owing to definition (1.3), that

$$\nabla \times E + \partial B/\partial t = 0, \tag{1.11}$$

which is the first of the first pair of the Maxwell equations (1.8). Applying $\nabla \times$ to the definition (1.3), one obtains

$$\nabla \times B = \nabla \times (\nabla \times A) = \nabla <\nabla, A> - \nabla^2 A =$$
$$= -\nabla(\partial \varphi/\partial t) - \partial^2 A/\partial t^2 + (\partial^2 A/\partial t^2 - \nabla^2 A) =$$
$$= \frac{\partial}{\partial t}(-\nabla \varphi - \partial A/\partial t) + J = \partial E/\partial t + J, \tag{1.12}$$

leading to
$$\nabla \times B = \partial E/\partial t + J,$$

which is the first of the second pair of the Maxwell equations (1.8). The final *"no magnetic charge"* equation

$$<\nabla, B> = <\nabla, \nabla \times A> = 0,$$

in (1.8) follows directly from the elementary identity $<\nabla, \nabla \times> = 0$, thereby completing the proof. $\square$

This proposition allows us to consider the electromagnetic potential vector-function $(\varphi, A) \in T^*(M^4)$ as a fundamental ingredient of the ambient *vacuum field medium*, by means of which we can try to describe the related physical behavior of charged point particles imbedded in space-time $M^4$. The following observation provides strong support for this approach:

**Observation.** *The Lorentz condition (1.4) actually means that the scalar potential field $\varphi : M^4 \to \mathbb{R}$ continuity relationship, whose origin lies in some new field conservation law, characterizes the deep intrinsic structure of the vacuum field medium.*

To make this observation more transparent and precise, let us recall the definition (56; 57; 70; 80) of the electric current $J : M^4 \to \mathbb{E}^3$ in the dynamical form

$$J := \rho v, \tag{1.13}$$

where the vector $v : M^4 \to \mathbb{E}^3$ is the corresponding charge velocity. Thus, the following continuity relationship

$$\partial \rho / \partial t + <\nabla, \rho v> = 0 \tag{1.14}$$

holds, which can easily be recast (123) as the integral conservation law

$$\frac{d}{dt} \int_{\Omega_t} \rho d^3 r = 0 \tag{1.15}$$

for the charge inside of any bounded domain $\Omega_t \subset \mathbb{E}^3$ moving in the space-time $M^4$ with respect to the natural evolution equation

$$dr/dt := v. \tag{1.16}$$

Following the above reasoning, we are led to the following result.

**Proposition 2.2.** *The Lorentz condition (1.4) is equivalent to the integral conservation law*

$$\frac{d}{dt} \int_{\Omega_t} \varphi d^3 r = 0, \tag{1.17}$$

*where $\Omega_t \subset \mathbb{E}^3$ is any bounded domain moving with respect to the evolution equation*

$$dr/dt := v, \tag{1.18}$$

*which represents the velocity vector of local potential field changes propagating in the Minkowski space-time $M^4$.*

*Proof.* Consider first the corresponding solutions to the potential field equations (1.5), taking into account condition (1.13). Owing to the results from (57; 70), one finds that

$$A = \varphi v, \tag{1.19}$$

which gives rise to the following form of the Lorentz condition (1.4):

$$\partial \varphi / \partial t + <\nabla, \varphi v> = 0. \tag{1.20}$$

This obviously can be rewritten (123) as the integral conservation law (1.17), so the proof is complete. □

The above proposition suggests a physically motivated interpretation of electrodynamic phenomena in terms of what should naturally be called *the vacuum potential field*, which determines the observable interactions between charged point particles. More precisely, we can *a priori* endow the ambient vacuum medium with a scalar potential field function $W := e\varphi : M^4 \to \mathbb{R}$, satisfying the governing *vacuum field equations*

$$\partial^2 W/\partial t^2 - \nabla^2 W = \rho, \quad \partial W/\partial t + <\nabla, Wv> = J, \tag{1.21}$$

taking into account that there are external sources besides material particles and possessing only a virtual capability for disturbing the vacuum field medium. Moreover, this vacuum potential field function $W : M^4 \to \mathbb{R}$ allows the natural potential energy interpretation, whose origin should be assigned not only to the charged interacting medium, but also to any other medium possessing interaction capabilities, including for instance, material particles interacting through the gravity.

This leads naturally to the next important step, which consists in deriving the equation governing the corresponding potential field $\bar{W} : M^4 \to \mathbb{R}$, assigned to the vacuum field medium in a neighborhood of any spatial point moving with velocity $u \in T(\mathbb{R}^3)$ and located at $R(t) \in \mathbb{E}^3$ at time $t \in \mathbb{R}$. As can be readily shown (53; 54), the corresponding evolution equation governing the related potential field function $\bar{W} : M^4 \to \mathbb{R}$ has the form

$$\frac{d}{dt}(-\bar{W}u) = -\nabla \bar{W}, \tag{1.22}$$

where $\bar{W} := W(t,r)|_{r \to R(t)}$, $u := dR(t)/dt$ at point particle location $(t, R(t)) \in M^4$.

Similarly, if there are two interacting point particles, located at points $R(t)$ and $R_f(t) \in \mathbb{E}^3$ at time $t \in \mathbb{R}$ and moving, respectively, with velocities $u := dR(t)/dt$ and $u_f := dR_f(t)/dt$, the corresponding potential field function $\bar{W} : M^4 \to \mathbb{R}$ for the particle located at point $R(t) \in \mathbb{E}^3$ should satisfy

$$\frac{d}{dt}[-\bar{W}(u - u_f)] = -\nabla \bar{W}. \tag{1.23}$$

The dynamical potential field equations (1.22) and (1.23) appear to have important properties and can be used as a means for representing classical electrodynamics. Consequently, we shall proceed to investigate their physical properties in more detail and compare them with classical results for Lorentz type forces arising in the electrodynamics of moving charged point particles in an external electromagnetic field.

In this investigation, we were strongly inspired by the works (81; 82; 89; 91; 93); especially by the interesting studies (87; 88) devoted to solving the classical problem of reconciling gravitational and electrodynamical charges within the Mach-Einstein ether paradigm. First, we revisit the classical Mach-Einstein relativistic electrodynamics of a moving charged point particle, and second, we study the resulting electrodynamic theories associated with our vacuum potential field dynamical equations (1.22) and (1.23), making use of the fundamental Lagrangian and Hamiltonian formalisms which were specially devised for this in (52; 55). The results obtained are used to apply the canonical Dirac quantization procedure to the corresponding energy conservation laws associated to the electrodynamic models considered.

## 2.2 Classical relativistic electrodynamics revisited

The classical relativistic electrodynamics of a freely moving charged point particle in the Minkowski space-time $M^4 := \mathbb{R} \times \mathbb{E}^3$ is based on the Lagrangian approach (56; 57; 70; 80) with Lagrangian function

$$\mathcal{L} := -m_0(1 - |u|^2)^{1/2}, \tag{1.24}$$

where $m_0 \in \mathbb{R}_+$ is the so-called particle rest mass and $u \in \mathbb{E}^3$ is its spatial velocity in the Euclidean space $\mathbb{E}^3$, expressed here and in the sequel in light speed units (with light speed $c$). The least action principle in the form

$$\delta S = 0, \quad S := -\int_{t_1}^{t_2} m_0(1 - |u|^2)^{1/2} dt \tag{1.25}$$

for any fixed temporal interval $[t_1, t_2] \subset \mathbb{R}$ gives rise to the well-known relativistic relationships for the mass of the particle

$$m = m_0(1 - |u|^2)^{-1/2}, \tag{1.26}$$

the momentum of the particle

$$p := mu = m_0 u(1 - |u|^2)^{-1/2} \tag{1.27}$$

and the energy of the particle

$$\mathcal{E}_0 = m = m_0(1 - |u|^2)^{-1/2}. \tag{1.28}$$

It follows from (57; 80), that the origin of the Lagrangian (1.24) can be extracted from the action

$$S := -\int_{t_1}^{t_2} m_0(1 - |u|^2)^{1/2} dt = -\int_{\tau_1}^{\tau_2} m_0 d\tau, \tag{1.29}$$

on the suitable temporal interval $[\tau_1, \tau_2] \subset \mathbb{R}$, where, by definition,

$$d\tau := dt(1 - |u|^2)^{1/2} \tag{1.30}$$

and $\tau \in \mathbb{R}$ is the so-called proper temporal parameter assigned to a freely moving particle with respect to the rest reference frame $\mathcal{K}_\tau$. The action (1.29) is rather questionable from the dynamical point of view, since it is physically defined with respect to the rest reference frame $\mathcal{K}_\tau$, giving rise to the constant action $S = -m_0(\tau_2 - \tau_1)$, as the limits of integrations $\tau_1 < \tau_2 \in \mathbb{R}$ were taken to be fixed from the very beginning. Moreover, considering this particle to have charge $e \in \mathbb{R}$ and be moving in the Minkowski space-time $M^4$ under action of an electromagnetic field $(\varphi, A) \in T^*(M^4)$, the corresponding classical (relativistic) action functional is chosen (see (52; 55–57; 70; 80)) as follows:

$$S := \int_{\tau_1}^{\tau_2} [-m_0 d\tau + e < A, \dot{r} > d\tau - e\varphi(1 - |u|^2)^{-1/2} d\tau], \tag{1.31}$$

with respect to the *rest reference frame*, parameterized by the Euclidean space-time variables $(\tau, r) \in \mathbb{E}^4$, where we have denoted $\dot{r} := dr/d\tau$ in contrast to the definition $u := dr/dt$. The action (1.31) can be rewritten with respect to the laboratory reference frame $\mathcal{K}$ moving with velocity vector $u \in T(\mathbb{R}^3)$ as

$$S = \int_{t_1}^{t_2} \mathcal{L} dt, \quad \mathcal{L} := -m_0(1 - |u|^2)^{1/2} + e <A, u> -e\varphi, \tag{1.32}$$

on the temporal interval $[t_1, t_2] \subset \mathbb{R}$, which gives rise to the following (56; 57; 70; 80) dynamical expressions

$$P = p + eA, \quad p = mu, \quad m = m_0(1 - |u|^2)^{-1/2}, \tag{1.33}$$

for the particle momentum and

$$\mathcal{E} = [m_0^2 + |P - eA|^2]^{1/2} + e\varphi \tag{1.34}$$

for the particle energy, where, by definition, $P \in T^*(\mathbb{R}^3)$ is the common momentum of the particle and the ambient electromagnetic field at a space-time point $(t, r) \in M^4$.

The expression (1.34) for the particle energy $\mathcal{E}_0$ also appears open to question, since the potential energy $e\varphi$, entering additively, has no affect on the particle mass $m = m_0(1 - |u|^2)^{-1/2}$. This was noticed by L. Brillouin (59), who remarked that since the potential energy has no affect on the particle mass, this tells us that "... any possibility of existence of a particle mass related with an external potential energy, is completely excluded". Moreover, it is necessary to stress here that the least action principle (1.32), formulated with respect to the laboratory reference frame $\mathcal{K}$ time parameter $t \in \mathbb{R}$, appears logically inadequate, for there is a strong physical inconsistency with other time parameters of the Lorentz equivalent reference frames. This was first mentioned by R. Feynman in (29), in his efforts to rewrite the Lorentz force expression with respect to the rest reference frame $\mathcal{K}_\tau$. This and other special relativity theory and electrodynamics problems induced many prominent physicists of the past (29; 59; 61; 64; 80) and present (4; 5; 60; 65; 66; 68; 69; 81; 82; 87; 89; 90; 93) to try to develop alternative relativity theories based on completely different space-time and matter structure principles.

There also is another controversial inference from the action expression (1.32). As one can easily show (56; 57; 70; 80), the corresponding dynamical equation for the Lorentz force is given as

$$dp/dt = F := eE + eu \times B. \tag{1.35}$$

We have defined here, as before,

$$E := -\partial A/\partial t - \nabla\varphi \tag{1.36}$$

for the corresponding electric field and

$$B := \nabla \times A \tag{1.37}$$

for the related magnetic field, acting on the charged point particle $e$. The expression (1.35) means, in particular, that the Lorentz force $F$ depends linearly on the particle velocity vector $u \in T(\mathbb{R}^3)$, and so there is a strong dependence on the reference frame with respect to which the charged particle $e$ moves. Attempts to reconcile this and some related controversies (29; 59;

60; 63) forced Einstein to devise his special relativity theory and proceed further to creating his general relativity theory trying to explain gravity by means of geometrization of space-time and matter in the Universe. Here we must mention that the classical Lagrangian function $\mathcal{L}$ in (1.32) is written in terms of a combination of terms expressed by means of both the Euclidean rest reference frame variables $(\tau, r) \in \mathbb{E}^4$ and arbitrarily chosen Minkowski reference frame variables $(t, r) \in M^4$.

These problems were recently analyzed using a completely different "no-geometry" approach (6; 53; 54; 60), where new dynamical equations were derived, which were free of the controversial elements mentioned above. Moreover, this approach avoided the introduction of the well-known Lorentz transformations of the space-time reference frames with respect to which the action functional (1.32) is invariant. From this point of view, there are interesting conclusions in (83) in which Galilean invariant Lagrangians possessing intrinsic Poincaré-Lorentz symmetry are reanalyzed. Next, we revisit the results obtained in (53; 54) from the classical Lagrangian and Hamiltonian formalisms (52) in order to shed new light on the physical underpinnings of the vacuum field theory approach to the investigation of combined electromagnetic and gravitational effects.

### 2.3 The vacuum field theory electrodynamics equations: Lagrangian analysis
#### 2.3.1 A point particle moving in a vacuum - an alternative electrodynamic model

In the vacuum field theory approach to combining electromagnetism and the gravity devised in (53; 54), the main vacuum potential field function $\bar{W} : M^4 \to \mathbb{R}$ related to a charged point particle $e$ satisfies the dynamical equation (1.21), namely

$$\frac{d}{dt}(-\bar{W}u) = -\nabla\bar{W} \tag{1.38}$$

in the case when the external charged particles are at rest, where, as above, $u := dr/dt$ is the particle velocity with respect to some reference frame .

To analyze the dynamical equation (1.38) from the Lagrangian point of view, we write the corresponding action functional as

$$S := -\int_{t_1}^{t_2} \bar{W} dt = -\int_{\tau_1}^{\tau_2} \bar{W}(1+\dot{r}^2)^{1/2} d\tau, \tag{1.39}$$

expressed with respect to the rest reference frame $\mathcal{K}_\tau$. Fixing the proper temporal parameters $\tau_1 < \tau_2 \in \mathbb{R}$, one finds from the least action principle ($\delta S = 0$) that

$$p := \partial \mathcal{L}/\partial \dot{r} = -\bar{W}\dot{r}(1+\dot{r}^2)^{-1/2} = -\bar{W}u,$$
$$\dot{p} := dp/d\tau = \partial \mathcal{L}/\partial r = -\nabla \bar{W}(1+\dot{r}^2)^{1/2}, \tag{1.40}$$

where, owing to (1.39), the corresponding Lagrangian function is

$$\mathcal{L} := -\bar{W}(1+\dot{r}^2)^{1/2}. \tag{1.41}$$

Recalling now the definition of the particle mass

$$m := -\bar{W} \tag{1.42}$$

and the relationships

$$d\tau = dt(1 - |u|^2)^{1/2}, \; \dot{r}d\tau = udt, \tag{1.43}$$

from (1.40) we easily obtain the dynamical equation (1.38). Moreover, one now readily finds that the dynamical mass, defined by means of expression (1.42), is given as

$$m = m_0(1 - |u|^2)^{-1/2}, \tag{1.44}$$

which coincides with the equation (1.26) of the preceding section. Now one can formulate the following proposition using the above results

**Proposition 2.3.** *The alternative freely moving point particle electrodynamic model (1.38) allows the least action formulation (1.39) with respect to the "rest" reference frame variables, where the Lagrangian function is given by expression (1.41). Its electrodynamics is completely equivalent to that of a classical relativistic freely moving point particle, described in Section 2.*

### 2.3.2 An interacting two charges system moving in a vacuum - an alternative vacuum-field theory electrodynamic model

We proceed now to the case when our charged point particle $e$ moves in the space-time with velocity vector $u \in T(\mathbb{R}^3)$ and interacts with another external charged point particle, moving with velocity vector $u_f = dr_f/dt \in T(\mathbb{R}^3)$ with respect to a laboratory reference frame $\mathcal{K}$, endowed with the coordinates $(t, r) \in M^4$. As was shown in (53; 54), the corresponding dynamical equation for the vacuum potential field function $\bar{W}' : M^4 \to \mathbb{R}$ with respect to the moving reference frame $\mathcal{K}'$, endowed with the corresponding coordinates $(t', r - r_f) \in M^4$, is given as

$$\frac{d}{dt'}[-\bar{W}'(u' - u'_f)] = -\nabla \bar{W}', \tag{1.45}$$

where $u'_f := dr_f/dt'$ and the potential function $\bar{W}' : M^4 \to \mathbb{R}$ is Lorentz related (57) with the laboratory $\bar{W} : M^4 \to \mathbb{R}$ potential function as

$$\bar{W}' = \bar{W}(1 - |u_f|^2)^{1/2}. \tag{1.46}$$

As the external charged particle moves in the space-time, it generates the related magnetic field $B := \nabla \times A$, whose magnetic vector potential $A : M^4 \to \mathbb{E}^3$ is defined, owing to the results of (53; 54; 60), as

$$eA' := \bar{W}'u'_f. \tag{1.47}$$

Whence, it easily follows from (1.40) that the particle momentum $p = -\bar{W}u$ equation (1.45) reduces to

$$\frac{d}{dt'}(p' + eA') = -\nabla \bar{W}', \tag{1.48}$$

if considered with respect to the moving reference frame $\mathcal{K}'$. The latter is equivalent, owing to the relationships

$$dt^2 = (dt')^2 + |dr_f|^2, \; (dt')^2 = d\tau^2 + |dr - dr_f|^2 \tag{1.49}$$

and definitions $p' := -\bar{W}'u' = -\bar{W}u = p \in \mathbb{E}^3$, $A' = \bar{W}'u'_f = \bar{W}u_f = A \in \mathbb{E}^3$ to the dynamical equation

$$\frac{d}{dt}(p + eA) = -\nabla \bar{W}(1 - |u_f|^2) \tag{1.50}$$

with respect to the laboratory reference frame $\mathcal{K}$. To represent the dynamical equation (1.50) in the classical Lagrangian formalism, we start from the following action functional, which naturally generalizes the functional (1.39):

$$S := -\int_{\tau_1}^{\tau_2} \bar{W}(1 + |\dot{r}' - \dot{\xi}'|^2)^{1/2} \, d\tau, \tag{1.51}$$

where $\dot{\xi}' = u'_f dt'/d\tau$, $d\tau = dt'(1 - |u' - u'_f|)^2)^{1/2}$, which takes into account the relative velocity of the charged point particle $e$ with respect to the reference frame $\mathcal{K}'$, moving with velocity $u_f \in T(\mathbb{R}^3)$ with respect to the reference frame $\mathcal{K}$. It is clear in this case that the charged point particle $e$ moves with velocity $u' - u'_f \in T(\mathbb{R}^3)$ with respect to the reference frame $\mathcal{K}'$ with respect to which the external charged particle is at rest.

Now we compute the least action variational condition $\delta S = 0$ taking into account that, owing to (1.51), the corresponding Lagrangian function is given as

$$\mathcal{L} := -\bar{W}'(1 + |\dot{r}' - \dot{\xi}'|^2)^{1/2}. \tag{1.52}$$

Hence, the common momentum of the particles is

$$\begin{aligned} P := \partial \mathcal{L}/\partial \dot{r}' &= -\bar{W}'(\dot{r}' - \dot{\xi}')(1 + |\dot{r}' - \dot{\xi}'|^2)^{-1/2} = \\ &= -\bar{W}'\dot{r}'(1 + |\dot{r}' - \dot{\xi}'|^2)^{-1/2} + \bar{W}'\dot{\xi}'(1 + |\dot{r}' - \dot{\xi}'|^2)^{-1/2} = \\ &= m'u' + eA' = p' + eA' = p + eA, \end{aligned} \tag{1.53}$$

and the dynamical equation is given as

$$\frac{d}{d\tau}(p' + eA') = -\nabla \bar{W}'(1 + |\dot{r} - \dot{r}_f|^2)^{1/2} \tag{1.54}$$

As $dt' = d\tau(1 + |\dot{r} - \dot{r}_f|^2)^{1/2}$ and $(1 + |\dot{r} - \dot{r}_f|^2)^{1/2} = (1 - |u' - u'_f|^2)^{-1/2}$, we obtain from (1.54) with respect to the reference frame $\mathcal{K}'$ the equation (1.48):

$$\frac{d}{dt'}(p' + eA') = -\nabla \bar{W}', \tag{1.55}$$

which finally reduces to the dynamical equation (1.50). Thus, the next proposition holds.

**Proposition 2.4.** *The alternative classical relativistic electrodynamic model (1.45) allows the least action formulation (1.51) with respect to the "rest" reference frame variables, where the Lagrangian function is given by expression (1.52).*

### 2.3.3 A moving charged point particle formulation dual to the classical alternative electrodynamic model

It is easy to see that the action functional (1.51) is written utilizing the classical Galilean transformations of reference frames. If we now consider the action functional (1.39) for a charged point particle moving with respect the reference frame $\mathcal{K}_\tau$, and take into account its interaction with an external magnetic field generated by the vector potential $A : M^4 \to \mathbb{E}^3$, it can be naturally generalized as

$$S := \int_{t_1}^{t_2} (-\bar{W}dt + e <A, dr>) = \int_{\tau_1}^{\tau_2} (-\bar{W}(1+\dot{r}^2)^{1/2} + e <A, \dot{r}>)d\tau, \qquad (1.56)$$

where $d\tau = dt(1-|u|^2)^{1/2}$.

Thus, the corresponding common particle-field momentum takes the form

$$P := \partial \mathcal{L}/\partial \dot{r} = -\bar{W}\dot{r}(1+\dot{r}^2)^{-1/2} + eA = \qquad (1.57)$$
$$= mu + eA := p + eA,$$

and satisfies

$$\dot{P} := dP/d\tau = \partial \mathcal{L}/\partial r = -\nabla \bar{W}(1+\dot{r}^2)^{1/2} + e\nabla <A, \dot{r}> = \qquad (1.58)$$
$$= -\nabla \bar{W}(1-|u|^2)^{-1/2} + e\nabla <A, u> (1-|u|^2)^{-1/2},$$

where

$$\mathcal{L} := -\bar{W}(1+\dot{r}^2)^{1/2} + e <A, \dot{r}> \qquad (1.59)$$

is the corresponding Lagrangian function. Since $d\tau = dt(1-|u|^2)^{1/2}$, one easily finds from (1.58) that

$$dP/dt = -\nabla \bar{W} + e\nabla <A, u>. \qquad (1.60)$$

Upon substituting (1.57) into (1.60) and making use of the well-known (57) identity

$$\nabla <a, b> = <a, \nabla> b + <b, \nabla> a + b \times (\nabla \times a) + a \times (\nabla \times b), \qquad (1.61)$$

where $a, b \in \mathbb{E}^3$ are arbitrary vector functions, we obtain the classical expression for the Lorentz force $F$ acting on the moving charged point particle $e$:

$$dp/dt := F = eE + eu \times B, \qquad (1.62)$$

where, by definition,

$$E := -e^{-1}\nabla \bar{W} - \partial A/\partial t \qquad (1.63)$$

is its associated electric field and

$$B := \nabla \times A \qquad (1.64)$$

is the corresponding magnetic field. This result can be summarized as follows:

**Proposition 2.5.** *The classical relativistic Lorentz force (1.62) allows the least action formulation (1.56) with respect to the rest reference frame variables, where the Lagrangian function is given by formula (1.59). Nonetheless, its electrodynamics, described by the Lorentz force (1.62), is not equivalent* to the classical relativistic moving point particle electrodynamics characterized by the Lorentz force (1.35) of Section 2, as the corresponding "inertial" mass relationships $m = -\bar{W}$ and $m = m_0(1 - |u|^2)^{-1/2}$ are strongly different.

As for the dynamical equation (1.54), it is easy to see that it is equivalent to

$$dp/dt = (-\nabla \bar{W} - edA/dt + e\nabla <A, u>) - e\nabla <A, u - u_f>, \qquad (1.65)$$

which, owing to (1.60) and (1.62), takes the following Lorentz type force form

$$dp/dt = eE + eu \times B - e\nabla <A, u - u_f>, \qquad (1.66)$$

which can be found in a slightly different form in (53; 54; 60).

Expressions (1.62) and (1.66) are equal to up to the gradient term $F_c := -e\nabla <A, u>$, which reconciles the Lorentz forces acting on a charged moving particle $e$ with respect to different reference frames. This fact is important for our vacuum field theory approach since it uses no special geometry and makes it possible to analyze both electromagnetic and gravitational fields simultaneously by employing the new definition of the dynamical mass by means of expression (1.42).

## 2.4 The vacuum field theory electrodynamics equations: Hamiltonian analysis

Any Lagrangian theory has an equivalent canonical Hamiltonian representation via the classical Legendre transformation(1; 2; 46; 56; 103). As we have already formulated our vacuum field theory of a moving charged particle $e$ in the Lagrangian form, we proceed now to its Hamiltonian analysis making use of the action functionals (1.39), (1.52) and (1.56).

Take, first, the Lagrangian function (1.41) and the momentum expression (1.40) for defining the corresponding Hamiltonian function

$$\begin{aligned} H &:= <p, \dot{r}> - \mathcal{L} = \\ &= - <p, p> \bar{W}^{-1}(1 - |p|^2/\bar{W}^2)^{-1/2} + \bar{W}(1 - |p|^2/\bar{W}^2)^{-1/2} = \\ &= -|p|^2 \bar{W}^{-1}(1 - |p|^2/\bar{W}^2)^{-1/2} + \bar{W}^2 \bar{W}^{-1}(1 - |p|^2/\bar{W}^2)^{-1/2} = \qquad (1.67) \\ &= -(\bar{W}^2 - |p|^2)(\bar{W}^2 - |p|^2)^{-1/2} = -(\bar{W}^2 - |p|^2)^{1/2}. \end{aligned}$$

Consequently, it is easy to show (1; 2; 56; 103) that the Hamiltonian function (1.67) is a conservation law of the dynamical field equation (1.38); that is, for all $\tau, t \in \mathbb{R}$

$$dH/dt = 0 = dH/d\tau, \qquad (1.68)$$

which naturally leads to an energy interpretation of $H$. Thus, we can represent the particle energy as

$$\mathcal{E} = (\bar{W}^2 - |p|^2)^{1/2}. \qquad (1.69)$$

Accordingly the Hamiltonian equivalent to the vacuum field equation (1.38) can be written as

$$\dot{r} := dr/d\tau = \partial H/\partial p = p(\bar{W}^2 - |p|^2)^{-1/2} \qquad (1.70)$$

$$\dot{p} := dp/d\tau = -\partial H/\partial r = \bar{W}\nabla\bar{W}(\bar{W}^2 - |p|^2)^{-1/2},$$

and we have the following result.

**Proposition 2.6.** *The alternative freely moving point particle electrodynamic model (1.38) allows the canonical Hamiltonian formulation (1.70) with respect to the "rest" reference frame variables, where the Hamiltonian function is given by expression (1.67). Its electrodynamics is completely equivalent to the classical relativistic freely moving point particle electrodynamics described in Section 2.*

In an analogous manner, one can now use the Lagrangian (1.52) to construct the Hamiltonian function for the dynamical field equation (1.50) describing with respect to the rest reference frame $\mathcal{K}_\tau$ the motion of charged particle $e$ in an external electromagnetic field in the canonical Hamiltonian form:

$$\dot{r} := dr/d\tau = \partial H/\partial P, \qquad \dot{P} := dP/d\tau = -\partial H/\partial r, \qquad (1.71)$$

where

$$\begin{aligned} H :=&< P, \dot{r} > -\mathcal{L} = \\ =&< P, \dot{r}_f - P\bar{W}'^{-1}(1 - |P|^2/\bar{W}'^2)^{-1/2} > +\bar{W}'[\bar{W}'^2(\bar{W}'^2 - |P|^2)^{-1}]^{1/2} = \\ =&< P, \dot{r}_f > +|P|^2(\bar{W}'^2 - |P|^2)^{-1/2} - \bar{W}'^2(\bar{W}'^2 - |P|^2)^{-1/2} = \\ =& -(\bar{W}'^2 - |P|^2)(\bar{W}'^2 - |P|^2)^{-1/2} + < P, \dot{r}_f > = \\ =& -(\bar{W}'^2 - |P|^2)^{1/2} - < eA, P > (\bar{W}'^2 - |P|^2)^{-1/2}. \end{aligned} \qquad (1.72)$$

Here we took into account that, owing to definitions (1.47) and (1.53),

$$\begin{aligned} e_f A' := \bar{W}' u'_f = \bar{W}' dr_f/dt' = \bar{W} u_f = e_f A = \\ = \bar{W}' \frac{dr_f}{d\tau} \cdot \frac{d\tau}{dt'} = \bar{W} \dot{r}_f (1 - |u' - u'_f|)^{1/2} = \\ = \bar{W}' \dot{r}_f (1 + |\dot{r} - \dot{r}_f|^2)^{-1/2} = \\ = -\bar{W}' \dot{r}_f (\bar{W}'^2 - |P|^2)^{1/2} \bar{W}'^{-1} = -\dot{r}_f (\bar{W}'^2 - |P|^2)^{1/2}, \end{aligned} \qquad (1.73)$$

or

$$\dot{r}_f = -e_f A \, (\bar{W}'^2 - |P|^2)^{-1/2}, \qquad (1.74)$$

where $A : \mathbb{M}^4 \to \mathbb{R}^3$ is the related magnetic vector potential generated by the moving external charged particle. Equations (1.72), owing to the relationship (1.74), can be rewritten with respect to the laboratory reference frame $\mathcal{K}$ in the form

$$dr/dt = u, \quad dp/dt = eE + eu \times B - e\nabla < A, u - u_f >, \qquad (1.75)$$

which coincides with the result (1.66).

Whence, we see that the Hamiltonian function (1.72) satisfies the energy conservation conditions

$$dH/dt = 0 = dH/d\tau, \qquad (1.76)$$

for all $\tau, t \in \mathbb{R}$, and that the suitable energy expression is

$$\mathcal{E} = (\bar{W}^2 - |p|^2)^{1/2} + e < A, P > (\bar{W}^2 - |p|^2)^{-1/2}, \qquad (1.77)$$

where the generalized momentum $P = p + eA$. The result (1.77) differs in an essential way from that obtained in (57), which makes use of the Einsteinian Lagrangian for a moving charged point particle $e$ in an external electromagnetic field. Thus, we obtain the following result:

**Proposition 2.7.** *The alternative classical relativistic electrodynamic model (1.75), which is intrinsically compatible with the classical Maxwell equations (1.8), allows the Hamiltonian formulation (1.71) with respect to the rest reference frame variables, where the Hamiltonian function is given by expression (1.72).*

The inference above is a natural candidate for experimental validation of our theory. It is strongly motivated by the following remark.

**Remark 2.8.** *It is necessary to mention here that the Lorentz force expression (1.75) uses the particle momentum $p = mu$, where the dynamical "mass" $m := -\bar{W}$ satisfies the energy conservation condition (1.77). The latter gives rise to the following crucial relationship between the particle energy $\mathcal{E}_0$ and its rest mass $m_0$ (at the velocity $u := 0$ at the initial time moment $t = 0 \in \mathbb{R}$) :*

$$\mathcal{E}_0 = m_0 \frac{(1 - |eA_0/m_0|^2)}{(1 - 2|eA_0/m_0|^2)^{1/2}}, \tag{1.78}$$

*or, equivalently, at the condition $|eA_0/m_0|^2 < 1/2$*

$$m_0 = \mathcal{E}_0 \left( \frac{1}{2} + |eA_0/\mathcal{E}_0|^2 \pm \frac{1}{2}\sqrt{1 - 4|eA_0/\mathcal{E}_0|^2} \right)^{1/2}, \tag{1.79}$$

*where $A_0 := A|_{t=0} \in \mathbb{E}^3$, which strongly differs from the classical expression $m_0 = \mathcal{E}_0 - e\varphi_0$, following from (1.34) and is not depending a priori on the external potential energy $e\varphi_0$. As the quantity $|eA_0/\mathcal{E}_0| \to 0$, the following asymptotical mass values follow from (1.79):*

$$m_0^{(+)} \simeq \mathcal{E}_0, \qquad m_0^{(-)} \simeq \pm\sqrt{2}|eA_0|. \tag{1.80}$$

*The first mass value $m_0^{(+)} \simeq \mathcal{E}_0$ is physically correct, giving rise to the bounded charged particle energy $\mathcal{E}_0$, but the second mass value $m_0^{(-)} \simeq \pm\sqrt{2}|eA_0|$ is not physical, as it gives rise to the vanishing denominator $(1 - 2|eA_0/m_0^{(-)}|^2)^{1/2} \simeq 0$ in (1.78), being equivalent to the unboundedness of the charged particle energy $\mathcal{E}_0$.*

To make this difference more clear, we now analyze the Lorentz force (1.62) from the Hamiltonian point of view based on the Lagrangian function (1.59). Thus, we obtain that the corresponding Hamiltonian function

$$\begin{aligned} H &:= <P, \dot{r}> - \mathcal{L} = <P, \dot{r}> + \bar{W}(1 + \dot{r}^2)^{1/2} - e<A, \dot{r}> = \\ &= <P - eA, \dot{r}> + \bar{W}(1 + \dot{r}^2)^{1/2} = \\ &= -<p, p>\bar{W}^{-1}(1 - |p|^2/\bar{W}^2)^{-1/2} + \bar{W}(1 - |p|^2/\bar{W}^2)^{-1/2} = \\ &= -(\bar{W}^2 - |p|^2)(\bar{W}^2 - |p|^2)^{-1/2} = -(\bar{W}^2 - |p|^2)^{1/2}. \end{aligned} \tag{1.81}$$

Since $p = P - eA$, expression (1.81) assumes the final *"no interaction"* (12; 57; 67; 80) form

$$H = -(\bar{W}^2 - |P - eA|^2)^{1/2}, \tag{1.82}$$

which is conserved with respect to the evolution equations (1.57) and (1.58), that is

$$dH/dt = 0 = dH/d\tau \tag{1.83}$$

for all $\tau, t \in \mathbb{R}$. These equations are equivalent with respect to the rest reference frame $\mathcal{K}_\tau$ to the following Hamiltonian system

$$\begin{aligned}\dot{r} &= \partial H/\partial P = (P - eA)(\bar{W}^2 - |P - eA|^2)^{-1/2}, \\ \dot{P} &= -\partial H/\partial r = (\bar{W}\nabla\bar{W} - \nabla <eA, (P - eA)>)(\bar{W}^2 - |P - eA|^2)^{-1/2},\end{aligned} \tag{1.84}$$

as one can readily check by direct calculations. Actually, the first equation

$$\begin{aligned}\dot{r} &= (P - eA)(\bar{W}^2 - |P - eA|^2)^{-1/2} = p(\bar{W}^2 - |p|^2)^{-1/2} = \\ &= mu(\bar{W}^2 - |p|^2)^{-1/2} = -\bar{W}u(\bar{W}^2 - |p|^2)^{-1/2} = u(1 - |u|^2)^{-1/2},\end{aligned} \tag{1.85}$$

holds, owing to the condition $d\tau = dt(1 - |u|^2)^{1/2}$ and definitions $p := mu$, $m = -\bar{W}$, postulated from the very beginning. Similarly we obtain that

$$\begin{aligned}\dot{P} &= -\nabla\bar{W}(1 - |p|^2/\bar{W}^2)^{-1/2} + \nabla <eA, u> (1 - |p|^2/\bar{W}^2)^{-1/2} = \\ &= -\nabla\bar{W}(1 - |u|^2)^{-1/2} + \nabla <eA, u> (1 - |u|^2)^{-1/2},\end{aligned} \tag{1.86}$$

coincides with equation (1.60)

$$dP/dt = -\nabla\bar{W} + e\nabla <A, u> \tag{1.87}$$

with respect to the evolution parameter $t \in \mathbb{R}$. This can be formulated as the next result.

**Proposition 2.9.** *The dual to the classical relativistic electrodynamic model (1.62) allows the canonical Hamiltonian formulation (1.84) with respect to the rest reference frame variables, where the Hamiltonian function is given by expression (1.82). Moreover, this formulation circumvents the "mass-potential energy" controversy associated with the classical electrodynamical model (1.32).*

The modified Lorentz force expression (1.62) and the related rest energy relationship are characterized by the following remark.

**Remark 2.10.** *If we make use of the modified relativistic Lorentz force expression (1.62) as an alternative to the classical one of (1.35), the corresponding particle energy expression (1.82) also gives rise to a different energy expression (at the velocity $u := 0 \in \mathbb{E}^3$ at the initial time $t = 0$) corresponding to the classical case (1.34); namely, $\mathcal{E}_0 = m_0$ instead of $\mathcal{E}_0 = m_0 + e\varphi_0$, where $\varphi_0 := \varphi|_{t=0}$.*

## 2.5 Concluding remarks

All of dynamical field equations discussed above are canonical Hamiltonian systems with respect to the corresponding proper rest reference frames $\mathcal{K}_\tau$, parameterized by suitable time parameters $\tau \in \mathbb{R}$. Upon passing to the basic laboratory reference frame $\mathcal{K}$ with the time parameter $t \in \mathbb{R}$, this naturally related Hamiltonian structure is lost, giving rise to a new interpretation of the real particle motion. Namely, one has an absolute sense to consider the relativistic particle dynamics only with respect to the proper rest reference frame, as otherwise this dynamics is completely relative and not uniquely defined with respect to all other reference frames. As for the Hamiltonian expressions (1.67), (1.72) and (1.82), one observes that they all depend strongly on the vacuum potential field function $\bar{W} : M^4 \to \mathbb{R}$, thereby avoiding the mass problem of the classical energy expression pointed out by L. Brillouin (59). It should be noted that the canonical Dirac quantization procedure can be applied only to the corresponding dynamical field systems considered with respect to their proper rest reference frames.

**Remark 2.11.** *Some comments are in order concerning the classical relativity principle. We have obtained our results without using the Lorentz transformations of reference frames - relying only on the natural notion of the rest reference frame and its suitable parametrization with respect to any other moving reference frames. It seems reasonable then that the true state changes of a moving charged particle e are exactly realized only with respect to its proper rest reference frame. Then the only remaining question would be about the physical justification of the corresponding relationship between time parameters of moving and rest reference frames.*

The relationship between reference frames that we have used through is expressed as

$$d\tau = dt(1 - |u|^2)^{1/2}, \quad (1.88)$$

where $u := dr/dt \in \mathbb{E}^3$ is the velocity with which the rest reference frame $\mathcal{K}_\tau$ moves with respect to another arbitrarily chosen reference frame $\mathcal{K}$. Expression (1.88) implies, in particular, that

$$dt^2 - dr^2 = d\tau^2, \quad (1.89)$$

which is identical to the classical infinitesimal Lorentz invariant. This is not a coincidence, since all our dynamical vacuum field equations were derived in turn (53; 54) from the governing equations of the vacuum potential field function $W : M^4 \to \mathbb{R}$ in the form

$$\partial^2 W/\partial t^2 - \nabla^2 W = \rho, \; \partial W/\partial t + \nabla(vW) = 0, \; \partial \rho/\partial t + <\nabla, J> = 0, \quad (1.90)$$

which are *a priori* Lorentz invariant. Here $\rho \in \mathbb{R}$ and $J = \rho v$ are, respectively, the charge and current densities, $v := dr/dt$ is the associated local velocity of the vacuum field medium evolution. Consequently, the dynamical infinitesimal Lorentz invariant (1.89) reflects this intrinsic structure of equations (1.90) with respect to the reference frame $\mathcal{K}_\tau$. Being rewritten in the nonstandard Euclidean form:

$$dt^2 = d\tau^2 + dr^2 \quad (1.91)$$

it gives rise to a completely different relationship between the reference frames $\mathcal{K}$ and $\mathcal{K}_\tau$, namely

$$dt = d\tau(1 + \dot{r}^2)^{1/2}, \quad (1.92)$$

where $\dot{r} := dr/d\tau$ is the related particle velocity with respect to the rest reference frame. Thus, we observe that all our Lagrangian analysis in Section 2 is based on the corresponding functional expressions written in these "Euclidean" space-time coordinates and with respect to which the least action principle was applied. So we see that there are two alternatives - the first is to apply the least action principle to the corresponding Lagrangian functions expressed in the Minkowski space-time variables with respect to an arbitrarily chosen reference frame $\mathcal{K}$, and the second is to apply the least action principle to the corresponding Lagrangian functions expressed in Euclidean space-time variables with respect to the rest reference frame $\mathcal{K}_\tau$.

This leads us to a slightly amusing but thought-provoking observation: It follows from our analysis that all of the results of classical special relativity related to the electrodynamics of charged point particles can be obtained (in a one-to-one correspondence) using our new definitions of the dynamical particle mass and the least action principle with respect to the associated Euclidean space-time variables in the rest reference frame .

An additional remark concerning the quantization procedure of the proposed electrodynamics models is in order: If the dynamical vacuum field equations are expressed in canonical Hamiltonian form, as we have done here, only straightforward technical details are required to quantize the equations and obtain the corresponding Schrödinger evolution equations in suitable Hilbert spaces of quantum states. There is another striking implication from our approach: the Einsteinian equivalence principle (29; 57; 63; 70; 80) is rendered superfluous for our vacuum field theory of electromagnetism and gravity.

Using the canonical Hamiltonian formalism devised here for the alternative charged point particle electrodynamics models, we found it rather easy to treat the Dirac quantization. The results obtained compared favorably with classical quantization, but it must be admitted that we still have not given a compelling physical motivation for our new models. This is something that we plan to revisit in future investigations. Another important aspect of our vacuum field theory no-geometry (geometry-free) approach to combining the electrodynamics with the gravity, is the manner in which it singles out the decisive role of the rest reference frame $\mathcal{K}_\tau$. More precisely, all of our electrodynamics models allow both the Lagrangian and Hamiltonian formulations with respect to the rest reference frame evolution parameter $\tau \in \mathbb{R}$, which are well suited the to canonical quantization. The physical nature of this fact still remains somewhat unclear. In fact, as far as we know (4; 5; 57; 63; 80), there is no physically reasonable explanation of this decisive role of the rest reference frame , except for that given by R. Feynman who argued in (70) that the relativistic expression for the classical Lorentz force (1.35) has physical sense only with respect to the rest reference frame variables $(\tau, r) \in \mathbb{E}^4$. In future research we plan to analyze the quantization scheme in more detail and begin work on formulating a vacuum quantum field theory of infinitely many particle systems.

## 3. The modified Lorentz force and the radiation theory

### 3.1 Introductory setting

Maxwell's equations may be represented by means of the electric and magnetic fields or by the electric and magnetic potentials. The latter were once considered as a purely mathematically motivated representation, having no physical significance.

The situation is actually not so simple now that evidence of the physical properties of the magnetic potential was demonstrated by Y. Aharonov and D. Bohm (92) in the formulation their "paradox" concerning the measurement of a magnetic field outside a separated region where it is vanishes. Later, similar effects were also revealed in the superconductivity theory of Josephson media. As the existence of any electromagnetic field in an ambient space can be tested only by its interaction with electric charges, the dynamics of the charged particles is very important. Charged particle dynamics was studied in detail by M. Faraday, A. Ampere and H. Lorentz using Newton's second law. These investigations led to the following representation for the Lorentz force

$$dp/dt = eE + e\frac{u}{c} \times B, \tag{1.93}$$

where $\mathbb{E}$ and $B \in \mathbb{E}^3$ are, respectively, electric and magnetic fields, acting on a point charged particle $e \in \mathbb{R}$ having momentum $p = mu$. Here $m \in \mathbb{R}_+$ is the particle mass and $u \in T(\mathbb{R}^3)$ is its velocity, measured with respect to a suitably chosen laboratory reference frame.

That the Lorentz force (1.93) is not completely correct was known to Lorentz. The defect can be seen from the nonuniform Maxwell equations for electromagnetic fields radiated by any accelerated charged particle, as easily seen from the well-known expressions for the Lienard-Wiechert potentials.

This fact inspired many physicists to "improve" the classical Lorentz force expression (1.93), and its modification was soon suggested by M. Abraham and P.A.M. Dirac, who found the so-called "radiation reaction" force induced by the self-interaction of a point charged particle:

$$\frac{dp}{dt} = eE + e\frac{u}{c} \times B - \frac{2e^2}{3e^3}\frac{d^2u}{dt^2}. \tag{1.94}$$

The additional force expression

$$F_s := -\frac{2e^2}{3c^3}\frac{d^2u}{dt^2}, \tag{1.95}$$

depending on the particle acceleration, immediately raised many questions concerning its physical meaning. For instance, a uniformly accelerated charged particle, owing to the expression (1.95), experiences no radiation reaction, contradicting the fact that any accelerated charged particle always radiates electromagnetic waves. This "paradox" was a challenging problem during the 20th century (73; 96–98; 100) and still has not been completely explained (101). As there exist different approaches to explanation this reaction radiation phenomenon, we mention here only some of the more popular ones such as the Wheeler-Feynman (99) "absorber radiation" theory, based on a very sophisticated elaboration of the retarded and advanced solutions to the nonuniform Maxwell equations, and Teitelbom's (95) approach which exploits the intrinsic structure of the electromagnetic energy tensor subject to the advanced and retarded solutions to the nonuniform Maxwell equations. It is also worth mentioning the very nontrivial development of Teitelbom's theory devised recently by (94) and applied to the non-abelian Yang-Mills equations, which naturally generalize the classical Maxwell equations.

**3.2 Radiation reaction force: the vacuum-field theory approach**

In the Section, we shall develop our vacuum field theory approach (6; 52–55) to the electromagnetic Maxwell and Lorentz theories in more detail and show that it is in complete agreement with the classical results. Moreover, it allows some nontrivial generalizations, which may have physical applications. For the radiation reaction force in the vacuum field theory approach, the modified Lorentz force, which was derived in Section 1, acting on a charged point particle $e$, is

$$dp/dt = -eE + e\frac{u}{c} \times B - \nabla < A, u - u_f >, \tag{1.96}$$

where $(\varphi, A) \in T^*(M^4)$ is the extended electromagnetic 4-vector potential. To take into account the self-interaction of this particle, we make use of the distributed charge density $\rho : M^4 \to \mathbb{R}$ satisfying the condition

$$e = \int_{\mathbb{R}^3} \rho(t,r) d^3 r \qquad (1.97)$$

for all $t \in \mathbb{R}$ with respect to a laboratory reference frame $\mathcal{K}$ with coordinates $(t,r) \in M^4$. Then, owing to 1.96 and results in (96), the self-interaction force can be expressed as

$$\begin{aligned} dp/dt := F_s = & -\frac{1}{c}\frac{d}{dt}\left[\int_{\mathbb{E}^3} d^3 r \rho(t,r) A_s(t,r)\right] - \\ & -\int_{\mathbb{E}^3} d^3 r \rho(t,r) \nabla \varphi_s(t,r) \left(1 - |u/c|^2\right) = \end{aligned} \qquad (1.98)$$

where we denoted by

$$\varphi_s(t,r) = \int_{\mathbb{E}^3} \frac{\rho(t',r')|_{ret} d^3 r'}{|r-r'|}, \quad A_s(t,r) = \frac{1}{c}\int_{\mathbb{E}^3} \frac{u(t')\rho(t',r')|_{ret} d^3 r'}{|r-r'|}, \qquad (1.99)$$

the well-known retarded Lienard-Wiechert potentials, which should be calculated at the retarded time parameter $t' := t - |r-r'|/c \in \mathbb{R}$. Taking additionally into account the continuity relationship

$$\partial \rho / \partial t + <\nabla, J> = 0 \qquad (1.100)$$

for the spatially distributed charge density $\rho : M^4 \to \mathbb{R}$ and current $J = \rho u : M^4 \to \mathbb{E}^3$ and the Taylor expansions for retarded potentials (1.99)

$$\begin{aligned} \varphi_s(t,r) &= \sum_{n \in \mathbb{Z}_+} \frac{\partial^n}{\partial t^n} \int_{\mathbb{E}^3} \frac{(-|r-r'|)^n}{c^n n!} \frac{\rho(t,r') d^3 r'}{|r-r'|}, \\ A_s(t,r) &= \sum_{n \in \mathbb{Z}_+} \frac{\partial^n}{\partial t^n} \int_{\mathbb{E}^3} \frac{(-|r-r'|)^n}{c^n n!} \frac{J(t,r') d^3 r'}{|r-r'|}, \end{aligned} \qquad (1.101)$$

from (1.98) and (1.101), assuming for brevity the spherical charge distribution, small enough value $|u|/c \ll 1$ and, respectively, slow acceleration, followed by calculations similar to those of (96; 124), one can obtain that

$$\begin{aligned} F_s = & \sum_{n \in \mathbb{Z}_+} \frac{(-1)^{n+1}}{n! c^n}(1 - |u/c|^2) \int_{\mathbb{E}^3} d^3 r \rho(t,r) \int_{\mathbb{E}^3} d^3 r' \frac{\partial^n}{\partial t^n} \rho(t,r') \nabla |r-r'|^{n-1} + \\ & + \frac{d}{dt}[\sum_{n \in \mathbb{Z}_+} \frac{(-1)^{n+1}}{n! c^n} \int_{\mathbb{E}^3} d^3 r \rho(t,r) \frac{|r-r'|^{n-1}}{c^2} \frac{\partial^n}{\partial t^n} J(t,r')] = \\ = & \sum_{m \in \mathbb{Z}_+} \frac{(-1)^{m+1}}{(m+2)! c^{m+2}}(1-|u/c|^2) \int_{\mathbb{E}^3} d^3 r \rho(t,r) \int_{\mathbb{E}^3} d^3 r' \frac{\partial^{m+2}}{\partial t^{m+2}} \rho(t,r') \nabla |r-r'|^{m+1} + \\ & + \frac{d}{dt}[\sum_{m \in \mathbb{Z}_+} \frac{(-1)^{m+1}}{m! c^m} \int_{\mathbb{E}^3} d^3 r \rho(t,r) \int_{\mathbb{E}^3} d^3 r' \frac{\partial^m}{\partial t^m}\left(\frac{|r-r'|^{m+1}}{c^2} J(t,r')\right)] = \end{aligned} \qquad (1.102)$$

$$= \sum_{m \in \mathbb{Z}_+} \frac{(-1)^{m+1}}{m! c^{m+2}} \int_{\mathbb{E}^3} d^3 r \rho(t,r)(1-|u/c|^2) \int_{\mathbb{E}^3} d^3 r' \frac{\partial^{m+1}}{\partial t^{m+1}} <J(t,r'),\nabla> |r-r'|^{m+1} =$$

$$+ \frac{d}{dt}[\sum_{m \in \mathbb{Z}_+} \frac{(-1)^{m+1}}{m! c^m} \int_{\mathbb{E}^3} d^3 r \rho(t,r) \int_{\mathbb{E}^3} d^3 r' \frac{|r-r'|^{m+1}}{c^2} \frac{\partial^m}{\partial t^m} (J(t,r'))] =$$

$$= \sum_{n \in \mathbb{Z}_+} \frac{(-1)^{n+1}}{n! c^{n+2}} (1-|u|^2) \int_{\mathbb{E}^3} d^3 r \rho(t,r) \int_{\mathbb{E}^3} d^3 r' |r-r'|^{n-1} \frac{\partial^{n+1}}{\partial t^{n+1}} [\frac{\partial \rho(t,r')}{\partial t} \frac{\nabla |r-r'|^{n+1}}{(n+2)(n+1)|r-r'|^{n-1}} + J(t,r')] +$$

$$+ \frac{d}{dt}[\sum_{m \in \mathbb{Z}_+} \frac{(-1)^{m+1}}{m! c^m} \int_{\mathbb{E}^3} d^3 r \rho(t,r) \int_{\mathbb{E}^3} d^3 r' \frac{|r-r'|^{m+1}}{c^2} \frac{\partial^m}{\partial t^m} J(t,r')] =$$

$$= \sum_{n \in \mathbb{Z}_+} \frac{(-1)^{n+1}}{n! c^{n+2}} (1-|u|^2) \int_{\mathbb{E}^3} d^3 r \rho(t,r)(1-|u|^2) \int_{\mathbb{E}^3} d^3 r' \frac{\partial^{n+1}}{\partial t^{n+1}} [-<\nabla, J(t,r')> \frac{|r-r'|^{n-1}(r-r')}{(n+2)}] +$$

$$+ \frac{d}{dt}[\sum_{m \in \mathbb{Z}_+} \frac{(-1)^{m+1}}{m! c^m} \int_{\mathbb{E}^3} d^3 r \rho(t,r) \int_{\mathbb{E}^3} d^3 r' \frac{|r-r'|^{m+1}}{c^2} \frac{\partial^m}{\partial t^m} J(t,r')] =$$

$$= \sum_{n \in \mathbb{Z}_+} \frac{(-1)^n}{n! c^{n+2}} (1-|u/c|^2) \int_{\mathbb{E}^3} d^3 r \rho(t,r) \int_{\mathbb{E}^3} d^3 r' |r-r'|^{n-1} \frac{\partial^{n+1}}{\partial t^{n+1}} \left( \frac{J(t,r')}{n+2} + \frac{n-1}{n+2} \frac{<r-r',J(t,r')>|}{|r-r'|^2} \right) +$$

$$+ \frac{d}{dt}[\sum_{m \in \mathbb{Z}_+} \frac{(-1)^{m+1}}{m! c^m} \int_{\mathbb{E}^3} d^3 r \rho(t,r) \int_{\mathbb{E}^3} d^3 r' \frac{|r-r'|^{m+1}}{c^2} \frac{\partial^m}{\partial t^m} J(t,r')].$$

The relationship above can be rewritten, owing to the charge continuity equation (1.100), gives rise to the radiation force expression

$$F_s = \sum_{n \in \mathbb{Z}_+} \frac{(-1)^n}{n! c^{n+2}} (1-|u/c|^2) \int_{\mathbb{E}^3} d^3 r \rho(t,r) \int_{\mathbb{E}^3} d^3 r' |r-r'|^{n-1} \frac{\partial^{n+1}}{\partial t^{n+1}} \left( \frac{J(t,r')}{n+2} + \frac{n-1}{n+2} \frac{<r-r',J(t,r')>(r-r')}{|r-r'|^2} \right) +$$

$$+ \frac{d}{dt}[\sum_{m \in \mathbb{Z}_+} \frac{(-1)^{m+1}}{m! c^m} \int_{\mathbb{E}^3} d^3 r \rho(t,r) \int_{\mathbb{E}^3} d^3 r' \frac{|r-r'|^{m+1}}{c^2} \frac{\partial^m}{\partial t^m} J(t,r')] =$$

$$= \sum_{n \in \mathbb{Z}_+} \frac{(-1)^{n+1}}{n! c^{n+2}} (1-|u/c|^2) \int_{\mathbb{E}^3} d^3 r \rho(t,r) \int_{\mathbb{E}^3} d^3 r' |r-r'|^{n-1} \frac{\partial^{n+1}}{\partial t^{n+1}} \left( \frac{J(t,r')}{n+2} + \frac{n-1}{n+2} \frac{|r-r',u|^2 J(t,r')}{|r-r'|^2 |u|^2} \right) +$$

$$+ \frac{d}{dt}[\sum_{m \in \mathbb{Z}_+} \frac{(-1)^{m+1}}{m! c^m} \int_{\mathbb{E}^3} d^3 r \rho(t,r) \int_{\mathbb{E}^3} d^3 r' \frac{|r-r'|^{m+1}}{c^2} \frac{\partial^m}{\partial t^m} J(t,r')].$$
(1.103)

Now, having applied to (1.103) the rotational symmetry property for calculation of the internal integral, one easily obtains that

$$F_s = \sum_{n \in \mathbb{Z}_+} \frac{(-1)^n}{n! c^{n+2}} (1-|u/c|^2) \int_{\mathbb{E}^3} d^3 r \rho(t,r) \int_{\mathbb{E}^3} d^3 r' |r-r'|^{n-1} \frac{\partial^{n+1}}{\partial t^{n+1}} \left( \frac{J(t,r')}{n+2} + \frac{(n-1)J(t,r')}{3(n+2)} \right) +$$

$$+ \frac{d}{dt}[\sum_{m \in \mathbb{Z}_+} \frac{(-1)^{m+1}}{m! c^m} \int_{\mathbb{E}^3} d^3 r \rho(t,r) \int_{\mathbb{E}^3} d^3 r' \frac{|r-r'|^{m+1}}{c^2} \frac{\partial^m}{\partial t^m} J(t,r')] =$$

$$= \sum_{n \in \mathbb{Z}_+} \frac{(-1)^n}{3n! c^{n+2}} (1-|u|^2) \int_{\mathbb{E}^3} d^3 r \rho(t,r) \int_{\mathbb{E}^3} d^3 r' |r-r'|^{n-1} \frac{\partial^{n+1}}{\partial t^{n+1}} J(t,r') +$$

$$+ \frac{d}{dt}[\sum_{m \in \mathbb{Z}_+} \frac{(-1)^{m+1}}{m! c^m} \int_{\mathbb{E}^3} d^3 r \rho(t,r) \int_{\mathbb{E}^3} d^3 r' \frac{|r-r'|^{m+1}}{c^2} \frac{\partial^m}{\partial t^m} J(t,r')] =$$

$$= \sum_{n\in\mathbb{Z}_+} \frac{(-1)^n}{3n!c^{n+2}}(1-|u/c|^2)\int_{\mathbb{E}^3} d^3r\rho(t,r)\int_{\mathbb{E}^3} d^3r'|r-r'|^{n-1}\frac{\partial^{n+1}}{\partial t^{n+1}}J(t,r')+$$

$$+\frac{d}{dt}[\sum_{m\in\mathbb{Z}_+} \frac{(-1)^{m+1}}{m!c^m}\int_{\mathbb{E}^3} d^3r\rho(t,r)\int_{\mathbb{E}^3} d^3r'\frac{|r-r'|^{m+1}}{c^2}\frac{\partial^m}{\partial t^m}J(t,r')] = \quad (1.104)$$

$$= \frac{d}{dt}[\sum_{m\in\mathbb{Z}_+} \frac{(-1)^{m+1}}{m!c^m}\int_{\mathbb{E}^3} d^3r\rho(t,r)\int_{\mathbb{E}^3} d^3r'\frac{|r-r'|^{m+1}}{c^2}\frac{\partial^m}{\partial t^m}[J(t,r')-\frac{1}{3}J(t,r')]-$$

$$-\sum_{n\in\mathbb{Z}_+} \frac{(-1)^m}{3m!c^{m+2}}(1-|u/c|^2)\int_{\mathbb{E}^3} d^3r\frac{\partial\rho(t,r)}{\partial t}\int_{\mathbb{E}^3} d^3r'|r-r'|^{m-1}\frac{\partial^{m+1}}{\partial t^{m+1}}J(t,r') =$$

$$= \frac{d}{dt}[\sum_{m\in\mathbb{Z}_+} \frac{(-1)^{m+1}}{m!c^m}\int_{\mathbb{E}^3} d^3r\rho(t,r)\int_{\mathbb{E}^3} d^3r'\frac{|r-r'|^{m+1}}{c^2}\frac{\partial^m}{\partial t^m}[J(t,r')-\frac{1}{3}J(t,r')]-$$

$$-\sum_{n\in\mathbb{Z}_+} \frac{(-1)^m}{3m!c^{m+2}}(1-|u/c|^2)\int_{\mathbb{E}^3} d^3r\frac{\partial\rho(t,r)}{\partial t}\int_{\mathbb{E}^3} d^3r'|r-r'|^{m-1}\frac{\partial^{m+1}}{\partial t^{m+1}}J(t,r') =$$

$$= \frac{d}{dt}[\sum_{m\in\mathbb{Z}_+} \frac{(-1)^{m+1}}{m!c^m}\int_{\mathbb{E}^3} d^3r\rho(t,r)\int_{\mathbb{E}^3} d^3r'\frac{|r-r'|^{m+1}}{c^2}\frac{\partial^m}{\partial t^m}[J(t,r')-\frac{1}{3}J(t,r')]+$$

$$+\sum_{n\in\mathbb{Z}_+} \frac{(-1)^m}{3m!c^{m+2}}(1-|u/c|^2)\int_{\mathbb{E}^3} d^3r <\nabla, J(t,r')> \int_{\mathbb{E}^3} d^3r'|r-r'|^{m-1}\frac{\partial^{m+1}}{\partial t^{m+1}}J(t,r') =$$

$$= \frac{d}{dt}[\sum_{m\in\mathbb{Z}_+} \frac{(-1)^{m+1}}{m!c^m}\int_{\mathbb{E}^3} d^3r\rho(t,r)\int_{\mathbb{E}^3} d^3r'\frac{|r-r'|^{m+1}}{c^2}\frac{\partial^m}{\partial t^m}[J(t,r')-\frac{1}{3}J(t,r')],$$

where we took into account (96) that in case of the spherical charge distribution the following equalities

$$\int_{\mathbb{E}^3} d^3r \int_{\mathbb{E}^3} d^3r'\rho(t,r)\rho(t,r')\frac{|<r-r',u(t)>|^2}{|r-r'|^2|u(t)|^2} = \frac{1}{3}e^2,$$

$$\int_{\mathbb{E}^3} d^3r <\nabla, J(t,r')> \int_{\mathbb{E}^3} d^3r'|r-r'|^{m-1}\frac{\partial^{m+1}}{\partial t^{m+1}}J(t,r') = 0, \quad (1.105)$$

$$\int_{\mathbb{E}^3} d^3r \int_{\mathbb{E}^3} d^3r\rho(t,r)\rho(t,r')\frac{(r-r')}{|r-r'|^3} = 0$$

hold. Thus, from (1.102) one easily finds up to the $O(1/c^4)$ accuracy the following radiation reaction force expression:

$$dp/dt = F_s = -\frac{d}{dt}\left(\frac{4\mathcal{E}_{es}}{3c^2}u(t)\right) - \frac{d}{dt}\left(\frac{2\mathcal{E}_{es}}{3c^2}|u/c|^2 u(t)\right) + \frac{2e^2}{3c^3}\frac{d^2u}{dt^2} + O(1/c^4) = \quad (1.106)$$

$$= -\frac{d}{dt}\left(\frac{4}{3}m_{0,es}u(t)(1+\frac{|u/c|^2}{2})\right) + \frac{2e^2}{3c^3}\frac{d^2u}{dt^2} + O(1/c^4) =$$

$$= -\frac{d}{dt}\left(\frac{4}{3}\frac{m_{0,es}u(t)}{(1-|u/c|^2)^{1/2}}\right) + \frac{2e^2}{3c^3}\frac{d^2u}{dt^2} + O(1/c^4) =$$

$$= -\frac{d}{dt}\left(\frac{4}{3}m_{es}u(t)\right) + \frac{2e^2}{3c^3}\frac{d^2u}{dt^2} + O(1/c^4),$$

where we defined, respectively, the electrostatic self-interaction repulsive energy as

$$\mathcal{E}_{es} := \frac{1}{2} \int_{\mathbb{R}^3} d^3r \int_{\mathbb{R}^3} d^3r' \frac{\rho(t,r)\rho(t,r')}{|r-r'|}, \tag{1.107}$$

the electromagnetic charged particle rest and inertial masses as

$$m_{0,es} := \frac{\mathcal{E}_{es}}{c^2}, \quad m_{es} := \frac{m_{0,es}}{(1-|u/c|^2)^{1/2}}. \tag{1.108}$$

Now from (1.96) one obtains that

$$\frac{d}{dt}\left[(m_g + \frac{4}{3}m_{es})u\right] = \frac{2e^2}{3c^3}\frac{d^2u}{dt^2} + O(1/c^4), \tag{1.109}$$

where we made use of the gravity inertial mass definition

$$m_g := -\bar{W}_g/c^2, \quad \nabla \bar{W}_g \simeq 0, \tag{1.110}$$

following from the vacuum field theory approach, where the $m_g \in \mathbb{R}_+$ is the corresponding gravitational mass of the charged particle $e$, generated by the gravity vacuum field potential $\bar{W}_g$. The corresponding radiation force

$$F_r = \frac{2e^2}{3c^3}\frac{d^2u}{dt^2} + O(1/c^4), \tag{1.111}$$

coinciding exactly with the classical Abraham-Lorentz-Dirac results. From (1.109) one follows that the observable physical charged particle mass consists of two impacts: the electromagnetic and gravitational components, giving rise to the final force expression

$$\frac{d}{dt}(m_{ph}u) = \frac{2e^2}{3c^3}\frac{d^2u}{dt^2} + O(1/c^4), \tag{1.112}$$

where

$$m_{ph} \simeq m_g + 4/3 m_{es}. \tag{1.113}$$

It means, in particular, that the real physically observed "inertial" mass $m_{ph}$ of a real electron strongly depends on the external physical interaction with the ambient vacuum medium, as it was recently demonstrated within completely different approaches in (132; 134), based on the vacuum Casimir effect considerations. Moreover, the assumed above boundedness of the electrostatic self-energy $\mathcal{E}_{es}$ appears to be completely equivalent to the existence of so-called intrinsic Poincaré type "tensions", analyzed in (112; 113; 134), and to the existence of a special compensating Coulomb "pressure", suggested in (132), guaranteeing the observable electron stability.

### 3.3 Conclusion

The charged particle radiation problem, revisited in this section, allows the explanation of the point charged particle mass as that of a compact and stable object, which should have a negative vacuum interaction potential $\bar{W} \in \mathbb{R}^3$ owing to (1.110). This negativity can be satisfied if and only if the quantity (1.110) is positive, thereby imposing certain nontrivial geometric constraints on the intrinsic charged particle structure (102). Moreover, as follows from the physically observed particle mass expressions (1.113), the electrostatic potential energy comprises the main portion of the full mass.

There exist different relativistic generalizations of the force expression (1.109), all of which suffer the same common physical inconsistency related to the no radiation effect of a charged point particle in uniform motion.

Another problem closely related to the radiation reaction force analyzed above is the search for an explanation to the Wheeler and Feynman reaction radiation mechanism, which is called the absorption radiation theory. This mechanism is strongly dependent upon the Mach type interaction of a charged point particle in an ambient vacuum electromagnetic medium. It is also interesting to observe some of the relationships between this problem and the one devised above in the context of the vacuum field theory approach, but more detailed and extended analyzes will be required to explain the connections.

## 4. Maxwell's equations and the Lorentz force derivation - the legacy of Feynman's approach

### 4.1 Poissonian analysis preliminaries

In 1948 R. Feynman presented but did not published (129; 130) a very interesting, in some respects "heretical", quantum-mechanical derivation of the classical Lorentz force acting on a charged particle under the influence of an external electromagnetic field. His result was analyzed by many authors (131; 133; 135–141) from different points of view, including its relativistic generalization (142). As this problem is completely classical, we reanalyze the Feynman's derivation from the classical Hamiltonian dynamics point of view on the coadjoint space $T^*(N), N \subset \mathbb{R}^3$, and construct its nontrivial generalization compatible with results (6; 52; 53) of Section 1, based on a recently devised vacuum field theory approach (52; 55). Upon obtaining the classical Maxwell electromagnetic equations, we supply the complete legacy of Feynman's approach to the Lorentz force and demonstrate its compatibility with the relativistic generalization presented in (52–55; 72).

Consider the motion of a charged point particle $e \in \mathbb{R}$ under the influence of an external electromagnetic field. For its description, following (115; 125; 126), it is convenient to introduce a trivial fiber bundle structure $\pi : M \to N$, $M = N \times G$, $N \subset \mathbb{R}^3$, with the abelian structure group $G := \mathbb{R} \backslash \{0\}$ equivariantly acting (1) on the canonically symplectic coadjoint space $T^*(M)$. Endow also this bundle with *a connection one-form* $\mathcal{A} : M \to T^*(M) \otimes \mathcal{G}$ defined as

$$\mathcal{A}(q;g) := g^{-1}(d+ <A(q),dq>)g \tag{1.114}$$

on the phase space $M$, where $q \in N$, $g \in G$ and $A : N \to T^*(N)$ is a differential form, constructed from the magnetic potential $A : N \to \mathbb{E}^3$. If $l : T^*(M) \to \mathcal{G}^*$ is the related momentum mapping, one can construct the reduced phase space $\bar{\mathcal{M}}_{\xi} := l^{-1}(e)/G \simeq T^*(N)$, where $e \in \mathcal{G}^* \simeq \mathbb{R}$ is taken to be fixed. This reduced space has the symplectic structure

$$\bar{\omega}_{\xi}^{(2)}(q,p) = <dp, \wedge dq> + ed <A(q), dq> . \tag{1.115}$$

From (1.115), one readily computes the respective reduced Poisson brackets on $T^*(N)$:

$$\{q^i, q^j\}_{\bar{\omega}_{\xi}^{(2)}} = 0, \quad \{p_j, q^i\}_{\bar{\omega}_{\xi}^{(2)}} = \delta^i_j, \quad \{p_i, p_j\}_{\bar{\omega}_{\xi}^{(2)}} = eF_{ji}(q) \tag{1.116}$$

for $i, j = \overline{1,3}$ with respect to the reference frame $\mathcal{K}(t,q)$, characterized by the phase space coordinates $(q, p) \in T^*(N)$. If one introduces a new momentum variable $\tilde{p} := p + eA(q)$ on $T^*(N)$, it is easy to verify that $\bar{\omega}^{(2)}_{\tilde{\zeta}} \to \tilde{\omega}^{(2)}_{\tilde{\zeta}} := <d\tilde{p}, \wedge dq>$, giving rise to the following "minimal coupling" canonical Poisson brackets (12; 125; 126):

$$\{q^i, q^j\}_{\tilde{\omega}^{(2)}_{\tilde{\zeta}}} = 0, \quad \{\tilde{p}_j, q^i\}_{\tilde{\omega}^{(2)}_{\tilde{\zeta}}} = \delta^i_j, \quad \{\tilde{p}_i, \tilde{p}_j\}_{\tilde{\omega}^{(2)}_{\tilde{\zeta}}} = 0 \tag{1.117}$$

for $i, j = \overline{1,3}$ with respect to the reference frame $\mathcal{K}_f(t, q - q_f)$, characterized by the phase space coordinates $(q, \tilde{p}) \in T^*(N)$, if and only if the Maxwell field equations

$$\partial F_{ij}/\partial q_k + \partial F_{jk}/\partial q_i + \partial F_{ki}/\partial q_j = 0 \tag{1.118}$$

are satisfied on $N$ for all $i, j, k = \overline{1,3}$ for the curvature tensor $F_{ij}(q) := \partial A_j/\partial q^i - \partial A_i/\partial q^j$, $i, j = \overline{1,3}, q \in N$.

### 4.2 The Lorentz type force and Maxwell electromagnetic field equations - the Lagrangian analysis

The Poisson structure (1.117) makes it possible to describe a charged particle $e \in \mathbb{R}$, located at point $q \in N \subset \mathbb{R}^3$, moving with a velocity $dq/dt := u \in T_q(N)$ with respect to the laboratory reference frame $\mathcal{K}(t, q)$, specified by coordinates $(t, q) \in M^4$, being under the electromagnetic influence of an external charged particle $e_f \in \mathbb{R}$ located at point $q_f \in N \subset \mathbb{R}^3$ and moving with respect to the same reference frame $\mathcal{K}(t, q)$ with a velocity $dq_f/dt := u_f \in T_{q_f}(N)$. Really, consider a new shifted reference frame $\mathcal{K}'_f(t', q - q_f)$ moving with respect to the reference frame $\mathcal{K}(t, q)$ with the velocity $u_f$. With respect to the reference frame $\mathcal{K}'_f(t', q - q_f)$, specified by coordinates $(t', q - q_f) \in M^4$, the charged point particle $e$ moves with the velocity $u' - u'_f := dr/dt' - dr_f/dt' \in T_{q-q_f}(N)$ and, respectively, the charged particle $e_f$ stays in rest. Then one can write down the standard *classical Lagrangian function* of the charged particle $e$ with a mass $m' \in \mathbb{R}_+$ subject to the reference frame $\mathcal{K}'_f(t', q - q_f)$:

$$\mathcal{L}_f(q, u') = \frac{m'}{2}|u' - u'_f|^2 - e\varphi', \tag{1.119}$$

and the suitably Lorentz transformed scalar potential $\varphi' = \varphi/(1 + |u'_f|^2) \in C^2(M^4; \mathbb{R})$ is the corresponding potential energy with respect to the reference frame $\mathcal{K}'_f(t', q - q_f)$. On the other hand, owing to (1.119) and the Poisson brackets (1.117) the following equality for the charged particle $e$ *canonical momentum* with respect to the reference frame $\mathcal{K}'_f(t', q - q_f)$ holds:

$$\tilde{p}' := p' + eA'(q) = \partial \mathcal{L}_f(q, u')/\partial u', \tag{1.120}$$

or, equivalently,

$$p' + eA'(q) = m'(u' - u'_f), \tag{1.121}$$

expressed in the units when the light speed $c = 1$. Taking into account that the charged particle $e$ momentum with respect to the reference frame $\mathcal{K}(t, q)$ equals $p' := m'u' \in T^*_q(N)$, one can easily obtain from (1.121) the important relationship

$$eA'(q) = -m'u'_f \tag{1.122}$$

for the vector potential $A \in C^2(M^4; \mathbb{E}^3)$, which was before obtained in (54; 55; 128) and described before in Section 1. Taking now into account (1.119) and (1.122) one finds the following Lagrangian equation:

$$\frac{d}{dt'}[p' + eA'(q)] = \partial \mathcal{L}_f(q, u')/\partial q = -e\nabla \varphi', \tag{1.123}$$

obtained before with respect to the shifted reference frame $\mathcal{K}'_f(t', q - q_f)$ in (54; 55) and giving rise, as the result of obvious relationships $p' = p, A' = A$, to the following charged point particle $e$ dynamics:

$$\begin{aligned}
dp/dt &= -e\partial A/\partial t - e\nabla\varphi(1 - |u_f|^2) - e<u, \nabla>A = \\
&= -e\partial A/\partial t - e\nabla\varphi - e<u, \nabla>A+ \\
&+ e\nabla<u, A> - e\nabla<A, u - u_f> = \\
&= -e(\partial A/\partial t + \nabla\varphi) + eu \times (\nabla \times A) - e\nabla<A, u - u_f>
\end{aligned} \tag{1.124}$$

with respect to the laboratory reference frame $\mathcal{K}(t, q)$. Based now on (1.124) we obtain the modified Lorentz type force

$$dp/dt = eE + eu \times B - e\nabla<u - u_f, A>, \tag{1.125}$$

where we put, as usually by definition,

$$E := -\partial A/\partial t - \nabla\varphi, \quad B := \nabla \times A, \tag{1.126}$$

and slightly differing from the classical (57; 70; 85; 96) Lorentz force expression

$$dp/dt = eE + eu \times B \tag{1.127}$$

by the gradient component

$$F_c := -e\nabla<A, u - u_f>. \tag{1.128}$$

Remark now that the modified Lorentz type force expression (1.125) can be naturally generalized to the relativistic case if to take into account that the standard Lorenz condition

$$\partial\varphi/\partial t + <\nabla, A> = 0 \tag{1.129}$$

is imposed on the electromagnetic potential $(\varphi, A) \in T^*(M^4)$.

Really, from (1.126) one obtains that the Lorentz invariant field equation

$$\partial^2 \varphi/\partial t^2 - \Delta\varphi = \rho_f, \tag{1.130}$$

where $\rho_f : M^4 \to \mathcal{D}'(M^4)$ is a generalized density function of the external charge distribution $e_f$. Following now by the calculations from (54; 55) we can easily find from (1.130) and the charge conservation law

$$\partial \rho_f / \partial t + <\nabla, J_f> = 0 \tag{1.131}$$

the next *Lorentz invariant* equation on the vector potential $A \in C^2(M^4; \mathbb{E}^3)$ :

$$\partial^2 A / \partial t^2 - \Delta A = J_f. \tag{1.132}$$

Moreover, relationships (1.126),(1.130) and (1.132) easily entail the true classical Maxwell equations

$$\nabla \times E = -\partial B / \partial t, \ \nabla \times B = \partial E / \partial t + J_f, \tag{1.133}$$
$$<\nabla, E> = \rho_f, \quad <\nabla, B> = 0$$

on the electromagnetic field $(E, B) \in C^2(M^4; \mathbb{E}^3 \times \mathbb{E}^3)$.

Consider now the Lorenz condition (1.129) and observe that it is equivalent to the following local conservation law:

$$\frac{d}{dt} \int_{\Omega_t} W d^3 q = 0, \tag{1.134}$$

giving rise to the important relationship for the magnetic potential $A \in C^2(M^4; \mathbb{E}^3)$

$$A = u_f \varphi \tag{1.135}$$

with respect to the laboratory reference frame $\mathcal{K}(t, q)$, where $\Omega_t \subset N$ is any open domain with the smooth boundary $\partial \Omega_t$, moving jointly with the charge distribution $e_f$ in the domain $N \subset \mathbb{R}^3$ with the corresponding velocity $u_f$. Taking into account relationship (1.122) one can find the expression for our charged particle $e$ "inertial" mass:

$$m = -\bar{W}, \quad \bar{W} := e\varphi, \tag{1.136}$$

coinciding with that obtained before in (54; 55; 128), where we denoted by $\bar{W} \in C^2(M^4; \mathbb{R})$ the corresponding potential energy of the charged point particle $e$.

### 4.3 The modified least action principle and the Hamiltonian analysis

Based on the representations (1.135) and (1.136) one can rewrite the determining Lagrangian equation (1.123) with respect to the shifted reference frame $\mathcal{K}'_f(t', q_f)$ as follows:

$$\frac{d}{dt'}[-\bar{W}'(u' - u'_f)] = -\nabla \bar{W}', \tag{1.137}$$

which is reduced to the Lorentz type force expression (1.125) calculated with respect to the laboratory reference frame $\mathcal{K}(t, q)$ :

$$dp/dt = eE + eu \times B - e\nabla <u - u_f, A>, \tag{1.138}$$

where we put, as before,

$$E := -\partial A / \partial t - \nabla \varphi, \quad B := \nabla \times A. \tag{1.139}$$

**Remark 4.1.** *It is interesting to remark here that equation (1.138) does not allow the Lagrangian representation with respect to the reference frame $\mathcal{K}(t,q)$ in contrast to that of equation (1.137) which is equivalent to (1.123).*

The remark above is a challenging source of our further analysis concerning the direct relativistic generalization of the modified Lorentz type force (1.125). Namely, the following proposition holds.

**Proposition 4.2.** *The Lorentz type force (1.125) in the case when the charged point particle e momentum is defined, owing to (1.136), as $p = -\bar{W}u$ is the exact relativistic expression allowing the Lagrangian representation of the charged particle e dynamics with respect to the rest reference frame $\mathcal{K}_\tau(\tau, q - q_f)$, related to the shifted reference frame $\mathcal{K}'_f(t', q - q_f)$ by means of the classical relativistic proper time infinitesimal transformation:*

$$dt' = d\tau(1 + |u' - u'_f|^2)^{1/2}, \qquad (1.140)$$

*where $\tau \in \mathbb{R}$ is the proper time parameter in the rest reference frame $\mathcal{K}_\tau(\tau, q - q_f)$.*

*Proof.* Take the following action functional with respect to the charged point particle $e$ rest reference frame $\mathcal{K}_\tau(\tau, q - q_f)$:

$$S^{(\tau)} := -\int_{t_1(\tau_1)}^{t_2(\tau_2)} \bar{W}' dt' = \int_{\tau_1}^{\tau_2} \bar{W}'(1 + |u' - u'_f|^2)^{1/2} d\tau, \qquad (1.141)$$

where the proper temporal values $\tau_1, \tau_2 \in \mathbb{R}$ are considered, in a Feynman spirit (70), to be fixed in contrast to the temporal parameters $t_2(\tau_2), t_2(\tau_2) \in \mathbb{R}$ depending, owing to (1.140), on the charged particle $e$ trajectory in the phase space $M^4$. The least action condition

$$\delta S^{(\tau)} = 0, \delta q(\tau_1) = 0 = \delta q(\tau_2), \qquad (1.142)$$

applied to (1.141), entails exactly the dynamical equation (1.137), being simultaneously equivalent to the relativistic Lorentz type force expression (1.125) with respect to the laboratory reference frame $\mathcal{K}(t, q)$. The latter proves the proposition. $\square$

Making use of the relationships between the reference frames $\mathcal{K}(t,q)$ and $\mathcal{K}_\tau(\tau, q - q_f)$ in the case when the external charge particle velocity $u_f = 0$, we can easily derive the following corollary.

**Corollary 4.3.** *Let the external charge point $e_f$ be in rest, that is the velocity $u_f = 0$. Then equation (1.137) reduces to*

$$\frac{d}{dt}(-\bar{W}u)] = -\nabla\bar{W}, \qquad (1.143)$$

*allowing the following conservation law:*

$$H_0 = \bar{W}(1 - |u|^2)^{1/2} = -(\bar{W}^2 - |p|^2)^{1/2}. \qquad (1.144)$$

*Moreover, equation (1.143) is Hamiltonian with respect to the canonical Poisson structure (1.117), Hamiltonian function (1.144) and the rest reference frame $\mathcal{K}_r(\tau, q)$:*

$$\left.\begin{array}{l} dq/d\tau := \partial H_0/\partial p = p(\bar{W}^2 - |p|^2)^{-1/2} \\ dp/d\tau := -\partial H_0/\partial q = -\bar{W}(\bar{W}^2 - |p|^2)^{-1/2}\nabla\bar{W} \end{array}\right\} \Rightarrow \left.\begin{array}{l} dq/dt = -p\bar{W}^{-1}, \\ dp/dt = -\nabla\bar{W} \end{array}\right\}. \qquad (1.145)$$

*In addition, if to define the rest particle mass $m_0 := -H_0|_{u=0}$, the "inertial" particle mass quantity $m \in \mathbb{R}$ obtains the well known classical relativistic form*

$$m = -\bar{W} = m_0(1 - |u|^2)^{-1/2}, \tag{1.146}$$

*depending on the particle velocity $u \in \mathbb{R}^3$.*

Concerning the general case of equation (1.137) analogous ones to the above results hold, which were also described in part in Section 1. We need only to mention that the induced Hamiltonian structure of the general equation (1.137) results naturally from its least action representation (1.141) and (1.142) with respect to the rest reference frame $\mathcal{K}_\tau(\tau, q)$.

### 4.4 Conclusion

We have demonstrated the complete legacy of the Feynman's approach to the Lorentz force based derivation of Maxwell's electromagnetic field equations. Moreover, we have succeeded in finding the exact relationship between Feynman's approach and the vacuum field approach devised in (54; 55). Thus, the results obtained provide deep physical backgrounds lying in the vacuum field theory approach. Consequently, one can simultaneously describe the origins of the physical phenomena of electromagnetic forces and gravity. Gravity is physically based on the particle *"inertial"* mass expression (1.136), which follows naturally from both the Feynman approach to the Lorentz type force derivation and the vacuum field approach.